\documentclass[12pt]{article}
\usepackage{amsmath,amssymb,amsthm,latexsym}
\usepackage{stmaryrd,wasysym,upgreek,mathrsfs,dsfont}
\usepackage[english]{babel}
\usepackage{graphicx,color}
\usepackage{xspace}
\usepackage[english]{babel}
\usepackage{pgf,pgfarrows,pgfnodes,pgfautomata,pgfheaps,pgfshade}
\usepackage{graphicx}
\usepackage{pifont,dsfont}
\usepackage{marvosym}
\usepackage{slashed}

\newcommand{\be}{\begin{equation}}
\newcommand{\ee}{\end{equation}}
\newcommand{\bqa}{\begin{eqnarray}}
\newcommand{\eqa}{\end{eqnarray}}
\newcommand{\bea}{\begin{eqnarray}}
\newcommand{\eea}{\end{eqnarray}}
\newcommand{\les}{\leqslant}

\def\lbt{\left(}
  \def\rbt{\right)}

\let\epsilon=\varepsilon

\newtheorem{definition}{Definition}

\begin{document}

\title{Towards Renormalizing Group Field Theory}
\author{Vincent Rivasseau\\
Laboratoire de Physique Th\'eorique, CNRS UMR 8627,\\
Universit\'e Paris XI,  F-91405 Orsay Cedex, France}

\maketitle
\begin{abstract}
We review some aspects of non commutative quantum field theory and group field theory,
in particular recent progress on the systematic study
of the scaling and renormalization properties of group field theory.
We thank G. Zoupanos and the organizers of the Corfu
2010 Workshop on Noncommutative Field Theory and Gravity  for encouraging us to write this review.
\end{abstract}

\begin{flushright}
LPT-20XX-xx
\end{flushright}
\medskip

\noindent  MSC: 81T08, Pacs numbers: 11.10.Cd, 11.10.Ef\\
\noindent  Key words: Group field theory, renormalization
expansion.

\section{Introduction}

Renormalization has been the soul and driving force of quantum field theory, from the early
computations of the Lamb shift and of the electron anomalous magnetic moment in 
quantum electrodynamics to the golden era of the 70's, when renormalization
of the non-Abelian gauge theories became the key to their successful adoption 
as theories of the electroweak and strong interactions.

At the same time Wilson and followers understood the true meaning of renormalization.
They explained it no longer as a mysterious hiding of infinities
under the rug but simply as the natural change of the laws 
of physics under change of the observation scale. 
This brilliant step immediately lead to progress in other domains of physics.
It became known under the (somewhat misleading)
name of the renormalization group.

In spite of these achievements, many physicists still dream 
of a fundamental theory of physics that would be free from any infinities
right from the start. It would have to include gravity, the ``rebellious interaction".
But since gravity is not perturbatively renormalizable in the naive sense, the mainstream
proposal is to include it into a framework such as string theory which includes new unobserved symmetries (in particular 
supersymmetry) and which would make every computation finite.

It is also often suggested that absence of infinities would mean that 
there is a fundamental scale, eg the Planck scale, beyond which 
we could probe nothing, not just because of temporary experimental limits,
but because nothing would exist. In this view physics and the flow of the renormalization
group would not exist beyond that scale.

But we might even prefer if quantum gravity after all 
was just like the rest of the standard model. In particular we would like if it has perturbative divergences
of a renormalizable type like electroweak and chromo quantum dynamics. These divergences could 
then drive an interesting renormalization group flow upstream to the Planck scale (or ``before" the big bang), 
just like the QCD flow drives the asymptotically free physics of quarks and gluons upstream to the hadronic scale.
These features could appear within a suitably enlarged and reinterpreted quantum field theory formalism. In order
to find such a formalism, we should adopt a guiding principle. We share with the loop quantum gravity community
the feeling that the most appealing such principle is background independence, the one principle that lead to the invention 
of general relativity in the first place \cite{Rov}. A most natural and promising quantum field theory formalism based on these criteria
is group field theory (GFT).
All these considerations lead us to propose the systematic
study of the scaling and renormalization properties of group field theory\footnote{This proposal was presented in some detail eg at 
the Nottingham conference in July 2008, and at the Beijing LQG 09 conference.}. 

The main encouragement which lead to such a program came from the ``happy end"
story of a simpler quantum field theory, 
the $\phi^{\star4 }_4$ theory on the non commutative Moyal space. For a while it was dubbed 
non renormalizable because
of the so-called ultraviolet-infrared mixing problem. Ultimately it turned out 
to become renormalizable and even asymptotically safe under a quite natural modification of the propagator \cite{GW,GW1,beta1}. 
This provides the scenario and main source 
of inspiration for our own views and hopes about quantum gravity.
We will recall some key points of this story in the next section,

We will subsequently review the modest but already promising results that have been obtained 
on group field theory  scaling and renormalization in the last two years.

\section{Scales}
 
In all quantum field theory models we know the renormalization group relies on three basic ingredients

\begin{itemize}

\item 
Scale decomposition,

\item 
Locality principle,

\item 
Power counting.

\end{itemize}

The theory is renormalizable when these three ingredients nicely fit together. Roughly speaking the key recipe is
to relate the divergences of the theory to particular subgraphs whose internal scales
are higher than their external scales. These graphs satisfy a locality principle. If by power counting their
local parts keep the form of the terms already present in the action of the model,
then the model is renormalizable. It means that the structure of the theory does not change with 
the observation scale, only the coupling constants move.

So right at the start lies the scale decomposition. It is perhaps both the most fundamental
and the most technical ingredient. A scale decomposition can indeed be realized in many  different technical ways, 
which should ultimately be equivalent. For instance a scale decomposition can be created through block spinning
or through the use of many different shapes of momentum cutoffs. At first sight scale seems forever linked to the notion of space and distance. Small 
scales correspond to the ultraviolet regime and to short distances; large scales correspond to the infrared regime
and to large distances. The names of infrared and ultraviolet come from Fourier analysis, and electrodynamics. Small details are
indeed detected by sending highly energetic probes that can read them thanks to their high momenta or short wave lengths.

However a background-independent quantum theory of space-time for gravity 
should start without assuming any fixed background metric, and ideally not even any particular space-time topology. 
So the first puzzling question arises:
how to define a background-independent notion of scale and a renormalization
group flow in the absence of any ordinary metric, and even of any particular topology?

The problem lies in our deeply rooted and often subconscious identification of
the notion of scale with that of distance. Fortunately we have some models at our disposal
to disentangle that identification. In the models of condensed matter scale is not given by ordinary distance
but by momentum distance to an extended singularity, the Fermi surface \cite{Riv1}. The 
Grosse-Wulkenhaar model is a
pedagogical toy model which allows us to experiment much further
with a renormalization group that mixes the usual notions of 
short and long distance scales.

Based on this experience in condensed matter and non commutative field theory we propose 
the following more abstract definition of scale:

\begin{definition} \label{defsca} A  {\emph{scale}}
is a {\emph{slice of eigenvalues of a propagator}} 
gently\footnote{The word {\emph{gently}} does not look scientific but should not be suppressed.
It means eg if the scale is defined through momentum cutoffs it is good to use smooth
cutoffs rather than sharp ones in order to ensure the corresponding dual decays of the sliced propagator.}
cut according to a  geometric progression.
\end{definition}

So we turned the problem into that of finding the right propagator
with the right non-trivial spectrum for quantum gravity. The progress lies in the fact
that such abstract propagators can exist and have non-trivial spectrum in absence of any space-time manifold.

\section{The Grosse-Wulkenhaar model}

The Grosse-Wulkenhaar $\phi^{\star 4}$ model on the Moyal space $R^4$ with 
$1/(p^2 + \Omega x^2)$ propagator is just renormalizable \cite{GW,GW1}. Its renormalization 
is very interesting, since all three main ingredients are subtly changed.
More precisely the renormalization of that model relies on

\begin{itemize}
 
\item A new scale decomposition which mixes the traditional ultraviolet and infrared scales,

\item A new locality principle, called Moyality,

\item A new power counting under which only regular\footnote{A {\emph{regular}} ribbon graph is a planar graph with all external legs on the outer boundary.} graphs do diverge.

\end{itemize}

A big unexpected bonus is the existence of a  {\emph {non trivial fixed point}} \cite{beta1} which makes that GW model 
essentially the prime and simplest candidate for a fully consistent four dimensional quantum field theory.

\subsection{The new scale decomposition}

The multiscale analysis relies on slicing the Grosse-Wulkenhaar propagator 
$1/[p^2 + \Omega^{2} (\tilde x )^2]$. This propagator has parametric representation in $x$ space 
\begin{eqnarray}
C(x,y)&=&\frac{\theta}{4 \Omega}
\Big(\frac{\Omega}{\pi \theta}\Big)^{}
\int_0^\infty \!\! d\alpha \;
e^{-\frac{\mu_{0}^2\theta}{4\Omega} \alpha}  \nonumber
\\ \nonumber
&& \hskip-1cm {  { \frac{1}{   (\sinh \alpha)^{2}}  
\exp \Big(-  \frac{\Omega }{\theta \sinh \alpha}  \|x-y\|^2
- \frac{\Omega}{\theta} \tanh \frac{\alpha}{2} (\|x\|^2{+}\|y\|^2)\Big)}}
\end{eqnarray}
  \hskip-.1cm
involving the  Mehler kernel rather than the heat kernel.

\medskip
A convenient slice decomposition is made through the parametric representation
\begin{align*}
C^{i}(x,y)= \int_{M^{-2i}}^{M^{-2(i-1)}} \!\! d\alpha \cdots
 &\les KM^{2i}e^{-c_{1} {{M^{2i}\|x-y\|^2}}-c_{2}{{M^{-2i}(\|x\|^2 + \|y\|^2)}}},       
\end{align*}
where $M$ is a fixed number, namely the ratio of the geometric slicing progression. 
The corresponding scales correspond to a mixture 
of the ordinary ultraviolet and infrared notions \cite{GW1}. 

\subsection{The Moyality principle}

\begin{figure}{\centerline{\includegraphics[width=3cm]{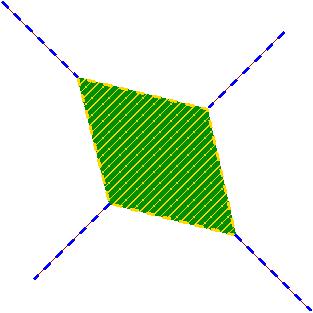}}}
\caption{The Moyal vertex}
\end{figure}

The Moyal vertex in direct space is proportional to
\bea \nonumber 
&&\int \prod_{i=1}^{4}d^{4}x^{i}\phi(x^{i})\, 
{{\delta(x_{1}-x_{2}+x_{3}-x_{4})}}
{\delta(x_{1}-x_{2}+x_{3}-x_{4})}
\hspace{.2cm}
\\&&
{{{{\exp\Big(
2\imath\theta^{-1}\lbt x_{1}\wedge x_{2}+x_{3}\wedge x_{4}\rbt\Big). }}}}
\eea

This vertex is {{\emph{non-local}}}
and oscillates. It has a
parallelogram shape, and the phase of the oscillation is proportional to its area.

\begin{figure}
{\includegraphics[width=6cm]{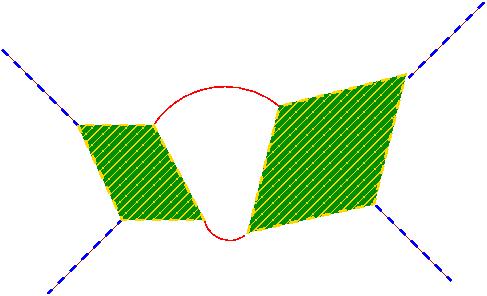}}\quad \quad
{\includegraphics[width=4cm]{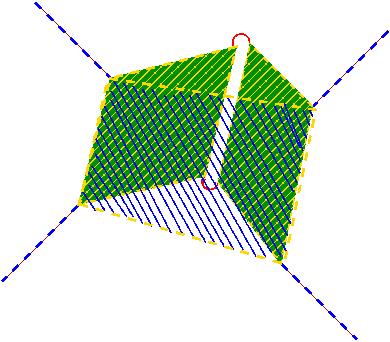}} 
\caption{The Moyality principle}\label{moyality}
\end{figure}

 In Figure \ref{moyality} it is shown how moving up the 
scales of the inner lines of a bubble glues its two parallelogram-shaped vertices
into what from the external legs point of view appears as a new parallelogram, and how the 
inner oscillations add up to the correct one for the new parallelogram, since the
area of the big parallelogram is the sum of the areas of the two smaller ones.

This Moyality principle applies only to regular, high subgraphs, where high means that
all internal scales are higher than all external scales.
Otherwise eg the oscillations won't add up correctly.
    
\subsection{Noncommutative Renormalization}

Renormalizability of the GW model combines the three elements. The new scale decomposition defines the ``high'' and ``low'' scales.
The new power counting tells us that only   {regular} graphs with two and four 
external legs must be renormalized.
The {{Moyality principle}} says that such ``high"   {regular} graphs 
look like Moyal products. The corresponding counterterms
are therefore of the form of the initial theory, thus the theory is {\emph {renormalizable}}.

An unexpected bonus is that it has a non-trivial ultraviolet fixed point \cite{beta1}.
Hence the theory should be fully consistent and non-trivial without cutoff. This 
is not expected to be the case of the ordinary $\phi^4_4$ theory, hence noncommutativity
has indeed improved renormalizability but in a very subtle sense.

See \cite{RivNC} for further references on noncommutative field theory.

\section{Group Field Theory}
 
Group field theory (GFT) \cite{LF,DO} lies at the crossroads between
the discretization of Cartan's first order formalism by loop 
quantum gravity, the higher dimensional tensorial extension 
of the matrix models relevant for 2D quantum gravity
and the ``Regge" or simplicial approach to quantum gravity

There are several enticing factors about GFT. The spin foams of loop gravity are exactly
the Feynman amplitudes of group field theory \cite{Rov}. However spin foam formalism alone still misses what we feel is a key ingredient,
namely how to weight correctly different topologies of space time. That's why the spin foam formalism alone is still an incomplete proposal to quantize gravity:
the amplitudes are there, but not the correct weights to sum them up.
In contrast, GFT is an attempt  to quantize completely 
space-time, summing {at the same time} over {metrics} and {topologies}. It 
assumes only the space-time dimension and
does not privilege any manifold at the start.  
It could provide a scenario for emergent large, manifold-like
classical space-time as a condensed phase  of 
more elementary space-time quanta.
If that is the case, there is no reason 
that some GFT's cannot hopefully be renormalized
through a suitably adapted multiscale RG. 

The main objections to GFT's that I heard during the last two years were
\begin{enumerate}
 
\item   The GFT's are tensor generalizations of matrix models; but whereas
matrix models really triangulate 2D Riemann surfaces, GFT's triangulate
much more singular objects (not even 
pseudo-manifolds with local singularities).

\item   There is no analog of the famous 1/N expansion for group field theories.

\item  The natural GFT's action does not look positive. Hence the
non-perturbative meaning of the theory is unclear.

\item  It seems difficult to identify ordinary space-time and physical observables in  GFT's.

\item  GFT's have infinities and it is not clear whether these infinities should or can 
be absorbed in a renormalization process;

\item GFT's don't predict space-time dimension, nor the standard
model.

\end{enumerate}

However at least some of the objections have been partly answered or are under detailed investigation.
The first two items have been more or less already solved through the invention of {\emph{colored}} group field theory \cite{Gur1,Gur2,Gur3,GurRiv}.
The third one may also be solved if, as colored field theory suggests, the GFT fields were eg Fermionic \cite{Gur1}.

Concerning the fourth item, one could say that to identify ordinary space-time and physical observables in  GFT's
is a non-trivial task, just as identifying hadrons out of quarks and gluons is. In particular the 
ultraspin limit $j \to \infty$ can be interpreted either as ultraviolet limit (on the group)
or as an infrared limit for a large ``chunk" of space-time. Quantum gravity
might therefore be a theory with some kind of ultraviolet/infrared mixing, more complicated than in the GW model.
However once large, topologically trivial 4d space time is obtained as an effective phase of the GFT,
general relativity on it should follow naturally
if we have respected the principle of background independence, 
which should then imply diffeomorphism invariance of the theory on any preferred specific manifold. 

Concerning the fifth  item it is exactly the point that our program would like to investigate in detail. 
Progress has been made essentially for ordinary topological
group field theories of the BF type and it is too early to draw conclusions for more general theories, 
that could well turn out to be just renormalizable
with a suitable choice of the propagator.

Finally it is true that GFT's don't predict space-time dimension, nor the standard
model, but the issue of completing GFT with matter fields is making progress \cite{matter1,Or1,Or2},
and it is not clear to us whether the competition really fares any better at this point.

\subsection{GFT as theories of Holonomies}

Consider a manifold $M$ with a connection $\Gamma$ (eg Levi-Civita of a metric).
The geometric information is {encoded} 
in the holonomies along all closed curves $\gamma$ in $M$.  

If $X$ is the vector field solution of 
the parallel transport equation
%\begin{figure}
%\includegraphics[width=3cm]{paralel.pdf}
%\end{figure}
\bea
 \frac{d\gamma^{\nu}}{dt} \partial_{\nu}X^{\mu}+\Gamma^{\mu}_{\nu\sigma}X^{\nu} \frac{d\gamma^{\sigma}}{dt}=0
\;, \quad X(0)= X_0 ,
\nonumber
\eea
then $X(T)=g X_0$ for some $g\in GL(TM_{\gamma(0)})$. $g$ (independent of $X_0$) is the holonomy along the curve $\gamma$. 
Suppose we {discretize} $M$ with flat $n$ dimensional simplices. Their boundary ($n-1$ dimensional) is also flat. The curvature is located at the ``joints'' of these blocks, that is 
at $n-2$ dimensional cells.
Hence holonomies $h$ are associated to blocks of codimension 2.

\subsection{$2D$ GFT as Matrix Models}

Consider a two dimensional surface $M$, and fix a triangulation of $M$.
The holonomy group of a surface is $G=U(1)$. To all vertices ({points}) in our triangulation 
we associate a holonomy $g$, encoding the deficit angle along any small curve encircling the vertex.
%\begin{figure}
 %\includegraphics[width=3cm]{discret.pdf}
%\end{figure}
The surface and its metric are specified by the gluing of the triangles and by the holonomies $g$. We associate 
to this surface a weight function $F(g,g',\dots)$.
Quantization of geometry should sum both over metrics compatible with a triangulation and over all triangulations.

The discrete information about the triangulation of the surface can be encoded in a ribbon ({2-stranded}) dual graph.
The ribbon vertices of the graph are dual to triangles and the ribbon lines are dual to edges.

%\begin{figure}
 %\includegraphics[width=2cm]{dualgr.pdf}
%\end{figure}
In the dual graph the group elements $g$ are associated to the sides of the ribbons, also called {strands}.
Suppose that the weight function $F$ factors into contributions of dual vertices and dual lines:
\bea
F=\prod_{V} V(g_1,g_2,g_3) \prod_{L}K(g_1,g_2) .
\nonumber
\eea
Then a 2-stranded graph is a ribbon Feynman graph (and its weight is the integrand 
of the Feynman amplitude) of a matrix model defined by the action 
\bea
&&
S=\frac{1}{2}\int_{G\times G} \phi(g_1,g_2) K^{-1}(g_1,g_2) \phi^*(g_1,g_2)\nonumber\\
&&+\lambda \int_{G\times G \times G} V(g_1,g_2,g_3)\phi(g_1,g_2)\phi(g_2,g_3)\phi(g_3,g_1) \; ,
 \nonumber
\eea
with $\phi(g_1,g_2)=\phi^*(g_2,g_1)$. Taking $K,V=1$ and developing $\phi$ in Fourier series
\bea
\phi(g_1,g_2)=\sum_{\mathbb{Z}\times \mathbb{Z}} e^{\imath m g_1} e^{-\imath n g_2}\phi_{mn}  \ , \nonumber
\eea
the action takes the more familiar form
\bea
S=\frac{1}{2} \sum_{mn}\phi_{mn}\phi^*_{mn} +\lambda 
\sum_{mnk} \phi_{mn} \phi_{n k} \phi_{k m} \ .
\nonumber
\eea

A complete correlation function of a matrix model 
\bea
 <\phi(g_1,g_2)\dots \phi(g_{2n-1},g_{2n})>
=\int [d\phi] e^{-S} { \phi(g_1,g_2)\dots \phi(g_{2n-1},g_{2n})} \; ,
\nonumber
\eea
admits the following interpretation.
The insertions $ \phi(g_1,g_2)$ fix a boundary triangulation and geometry.
Each Feynman graph fixes a bulk triangulation.
The amplitude of a graph is the sum over all bulk geometries compatible with the triangulation.
The correlation function automatically sums the weights of all triangulations and geometries
compatible with the boundary! As such the matrix models provide a quantization 
of 2D {\it {pure}} gravity.

\subsection{General Relativity in three dimensions}

In Cartan's formalism, 3D quantum gravity is expressed in terms of a dreibein 
\bea  {{e}}^i (x) = {{e}}^i_a (x) dx^a \nonumber
\eea
and a spin connection $\omega$ with values
in the $so(3)$ Lie algebra
\bea  \omega^i (x) = \omega^i_a (x) dx^a . \nonumber
\eea
The action is 
\bea  S({{e}},\omega) = \int {{e}}_i \wedge  F(\omega)^i \nonumber
\eea
where $F$ is the curvature of $\omega:$ 
\bea  F(\omega)  = d \omega + \omega \wedge \omega . \nonumber
\eea
 
Varying $\omega$ gives Cartan's equation $De=0$. Varying 
${e}$ gives $F=0$, hence  a {\emph{flat}} space-time. 
3D gravity is a {topological theory} with only global observables and no 
propagating degrees of freedom. 
Nevertheless the theory is physically interesting. Matter can be added;
point particles do not curve space but induce an angular deficit proportional to their mass.
Therefore in 3D gravity there is a limit ($2 \pi$) to any point-particle's mass. 

\subsection{Group Field Theory in three dimensions}

In three dimensions we must use the holonomy group $G=SU(2)$, the universal covering of the
$SO(3)$ rotations group.
Group elements are associated to edges in the triangulation (codimension 2).
Each field $\phi$ is associated to a triangular face of a tetrahedron, therefore it has three arguments.
The propagator $K$ is the inverse of the quadratic part
$$
\frac{1}{2}\int_{G^{3}} \phi(g_a,g_b,g_c) K^{-1}(g_a,g_b,g_c) \phi^*(g_a,g_b,g_c).
$$
This vertex is dual to a tetrahedron. A tetrahedron is bounded by four triangles therefore the vertex is a $\phi^4$ term, namely
$$
\lambda \int_{G^{6}}V(g,\dots, g)\phi(g_{03},g_{02},g_{01})
\phi(g_{01},g_{13},g_{12})\phi(g_{12},g_{02},g_{23})
\phi(g_{23},g_{13},g_{03}).
$$
What characterizes vertices in QFT is some kind of locality property. Similarly we propose

 \begin{definition} 
A vertex joining $2p$ strands in GFT is called {\emph{simple}} if it has for kernel in direct group space
{  {a product of 
$p$ delta functions}} matching strands two by two in different half-lines.
\end{definition}

The natural 3D GFT tetrahedron vertex in 3 dimensions is simple in this sense (with $p=6$) as it writes 
\bea
V[\phi]&=&\lambda \int \left(\prod_{i=1}^{12} d g_i\right)\phi(g_1,g_2,g_3) \phi(g_4,g_5,g_6) \nonumber 
\\ &&\ Ê\phi(g_7,g_8,g_9)\phi(g_{10},g_{11},g_{12})
Q(g_1, .. g_{12}),   \label{vertexdirect}  \nonumber
\eea
with a kernel 
\be  Q(g_1, .. g_{12}) = \delta(g_3g_4^{-1})\delta(g_2g_8^{-1})\delta(g_6g_7^{-1})\delta(g_9g_{10}^{-1})\delta(g_5g_{11}^{-1})\delta(g_1g_{12}^{-1}) \nonumber .
\ee

The 3D GFT propagator should implement the {\emph{flatness condition}}. The propagator 
\be
[C \phi] (g_1, g_2, g_3)  = \int dh \phi (hg_1, hg_2, hg_3)\nonumber
\ee 
is a projector onto gauge invariant fields $\phi (hg, hg', hg")= \phi (g, g', g")$. It has kernel
\be C =  \int dh  \, \delta (g_1 h {g'}_1^{-1}) \, \delta(g_2 h {g'}_2^{-1}) \,
\delta(g_3 h {g'}_3^{-1}) \nonumber .
\ee
Performing the integrations over all group elements of the vertices and keeping
the $h$ elements unintegrated
gives for each triangulation, or 3-stranded graph $G$ a Feynman amplitude  
\be  Z_G = \int \prod_e dh_e \prod_f \delta(g _f)\nonumber
\ee
where $g_f = \vec \prod_{e \in f} h_e$ is the holonomy along the face $f$.
Therefore 3D GFT with the projector $C$ as propagator 
implements exactly the correct flatness conditions of 3D gravity \cite{Boul}!
Fourier transforming the model, one gets Ponzano-Regge amplitudes.

This 3D GFT  is topological and the amplitude of a graph changes through a global multiplicative 
factor under the so-called Pachner moves.
Amplitudes may be infinite, for instance the tetrahedron graph or complete graph $K_4$
diverges as $\Lambda^3$
at zero external data, if $\Lambda$ is the ultraviolet cutoff on the size of $j$'s.
Regularization by going to a quantum group at a root of unity leads to well-defined
topological invariants of the triangulation, namely the Turaev-Viro invariants.
However the theory seems unsuited for a true RG analysis, as the propagator has spectrum
limited to 0 and 1. It cannot therefore be used to define truly scales in the abstract sense of Definition \ref{defsca}.
  
\subsection{The 4 Dimensional Case}
 
In the first order Cartan formalism, the action is proportional to $\int {}^\star [{e} \wedge {{e}} ]\wedge F$
where again the vierbein ${{e}}$ and the spin connection $\omega$ are considered
independent variables.

An approach to gravity in 4 dimensions is to consider it as a constrained $BF$ theory. The 
$BF$ theory, of action $\int B\wedge F$, involves an arbitrary two form $B$.
As not all two-forms $B$ in four dimenions are of the form
$e \wedge e$, one needs to implement a certain number of {\emph{constraints}} on the $B$ field,
the so called Plebanski constraints.
These constraints render 4D gravity much more complicated
and interesting since they are responsible for the local propagating degrees of freedom,
the gravitational waves, as constraints on $B$ allow richer set of $F$'s than just $F=0$.

Still another action classically equivalent to the Einstein-Hilbert
action is the Holst action:
\be
S = -\frac{1}{8\pi G}  \int  [ {}^\star(e \wedge e) + \frac{1}{ \gamma } (e \wedge e)]\wedge F  ,
\nonumber\ee 
where the first term is the Palatini action, and the second one (the Holst term) is topological. 
The parameter $\gamma$ (called the Barbero-Immirzi parameter) plays no r\^ole  
at the classical level but is crucial to the loop quantization of
4D gravity \cite{Rov}. We can then rewrite this action \`a la Plebanski (i.e. $BF$ theory plus constraints)
and get the Palatini-Holst-Plebanski functional integral measure, loosely written as:
\be
d\nu = 
\frac{1}{ Z} e^{ -\frac{1}{8\pi G}  \int  [ {}^\star B + \frac{1}{ \gamma } B ]\wedge F }  \delta( B = e\wedge e)
\; D B \; D \omega . \nonumber\ee

\subsubsection{4D GFT Vertex}

The vertex is the easy part as it  should be again given by gluing rules for the five tetrahedra which join into
{pentachores}.

\begin{figure}
\begin{center}
\includegraphics[scale=0.15,angle=-90]{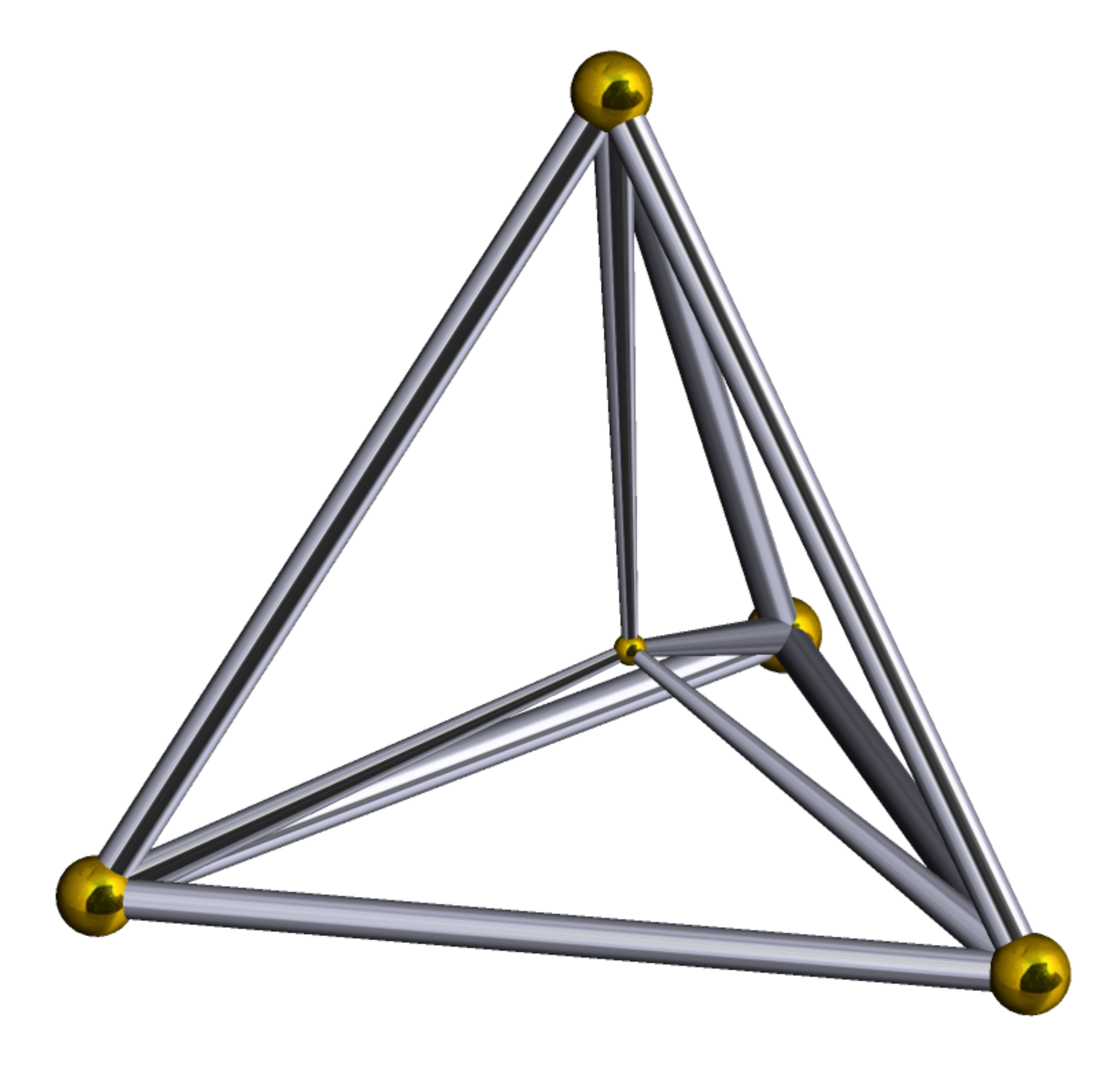}
\caption{The pentachore}
\end{center}
\end{figure}
The corresponding graphs  are made of vertices dual to the pentachores, of four-stranded edges dual to tetrahedra, and of faces or closed thread circuits dual to triangles.
If we keep the same propagator than in the Boulatov theory   
\be
[C \phi] (g_1, g_2, g_3, g_4)  = \int dh \phi (hg_1, hg_2, hg_3, hg_4) ,
\ee
we get a $\phi^5$ GFT with 4-stranded graphs called the Ooguri group field theory \cite{Oog}. It is not 
4D gravity, but a discretization of the 4D BF theory, ie the Plebanski constraints are missing.
Again such a topological version of gravity is unsuited for RG analysis.

A first  attempt to implement the 
Plebanski constraints in this language is due to Barrett and Crane. They suggested 
one should restrict to   {simple} representations of $SU(2) \times SU(2)$, namely those satisfying
$j_+ = j_-$. 
However this does not seem to work because the Barrett-Crane proposal
may implement the Plebanski constraints too strongly. In particular 
it may not lead to the correct angular degrees of freedom of the graviton.
  
\section{The EPR/FK theory}
 
Parallel works by Engle-Pereira-Rovelli and by Freidel-Krasnov, with contributions of Livine and Speziale, lead in 2007 to an improved spin foam model \cite{EPR,FK,LS}
which may better incorporate the Plebanski constraints and the angular degrees of freedom of the graviton. It
translates into an new proposal for a 4D GFT \cite{BGR} with an
improved   {propagator} which incorporates 

\begin{itemize} 
\item
A $SU(2) \times SU(2)$ averaging  $C$ at both ends like in Ooguri theory

\item
A  simplicity constraint which depends on the value of $\gamma$:
\bea  j_+ = \frac{1+\gamma}{2} j \;\; ;  \;\;\;\;  j_- = \frac{1-\gamma}{2} j   \;\; {\rm for} \;\; 0< \gamma <1 .
\eea

\item
{A new  projector $S$} averaging on a   {single $SU(2)$} residual gauge invariance in the
middle of the propagator.

\end{itemize}

\subsection{Renormalization group, at last?}

The main new feature of the EPR/FK models is that the full propagator can therefore be written as
$K = CSC$ with $C^2 = C$ and $S^2 = S$, hence $C$ and $S$
are two projectors, {but they do not commute},
hence $K$ has {non-trivial spectrum!}
This model is therefore a natural starting point for a RG analysis along our program.

A first step in this direction has been performed in 2008 \cite{PRS}. In some normalization of the theory
a tantalizing logarithmic divergence
for a radiative correction to the coupling constant and a
$\Lambda^6$ divergence for the graph $G_2$

\centerline{\includegraphics[width=3.5cm]{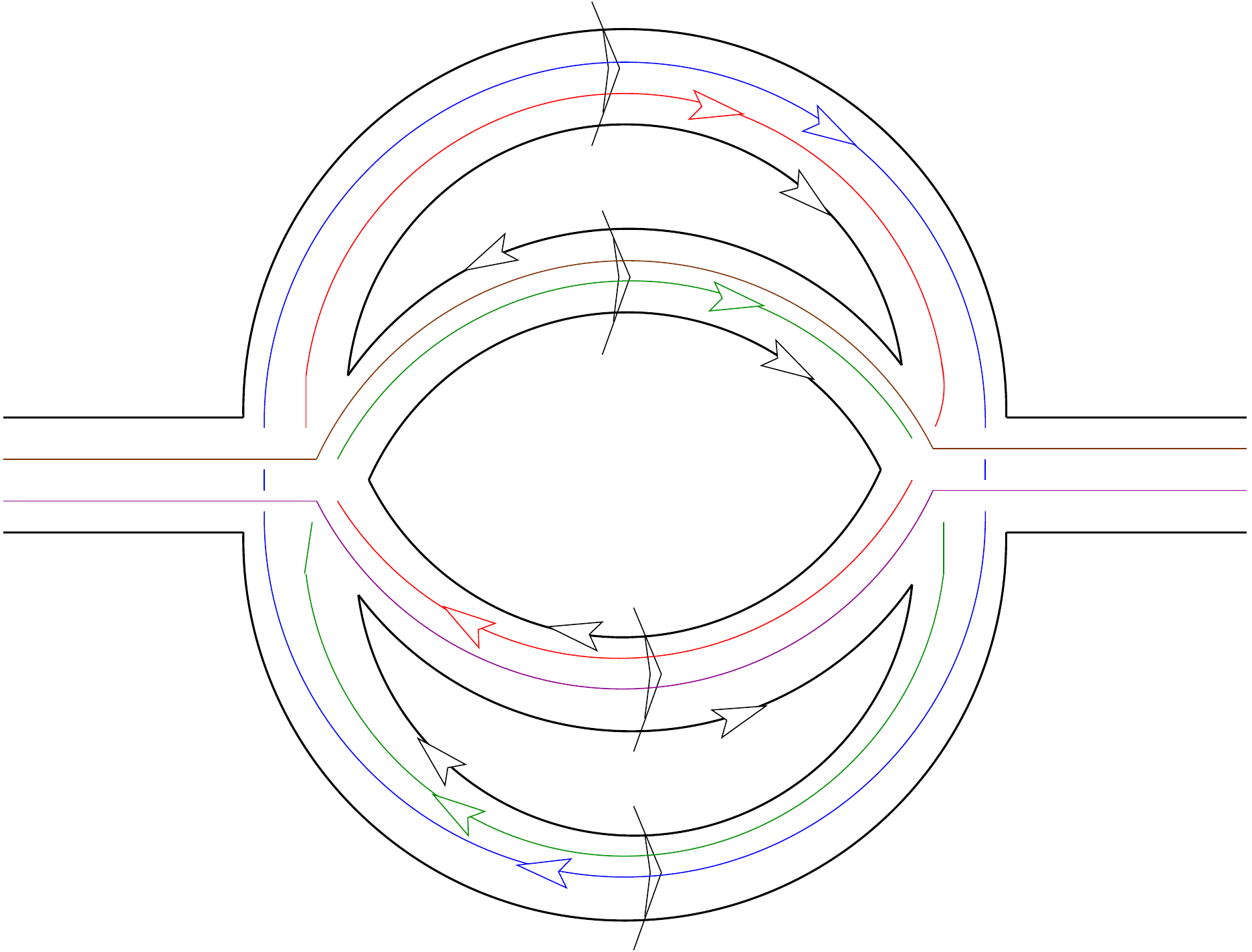}}
have been found.

\subsection{What has been done so far}
\footnote{This section mostly summarizes joint work with J. Ben Geloun, R. Gurau, T. Krajewski, J. Magnen,  K. Noui, 
M. Smerlak,  A. Tanasa and P. Vitale.}
We have investigated the power counting of the Boulatov model,
and found the first uniform bounds for its Freidel-Louapre constructive regularization,
using the loop vertex expansion technique \cite{MNRS}.

We have computed in the Abelian case
the power counting of graphs for $BF$ theory, and found how to relate them to the
number of bubbles of the graphs in the colored case, along the lines of \cite{FGO,Gur1}.
The Abelian amplitudes lead to new class of topological polynomials for stranded graphs
\cite{BMR, BKMR}.
See also the related work of Bonzom and Smerlak \cite{BS1,BS2} and \cite{Gur5} for colored polynomials.
  
We have started the study of the general superficial degree of divergence
of EPR/FK graphs through saddle point analysis using coherent states
techniques \cite{KMRTV}. We recover in many cases the Abelian counting, and find also
corrections to it in general.
But saddle points may come in many varieties, degenerate or not. This richness is
interesting but a major challenge.
We recovered the $\Lambda^6$ divergence of the EPR/FK
 graph $G_2$ in the case of non-degenerate saddle point configurations, but also
 a $\Lambda^9$ divergence for the maximally degenerate saddle points \cite{KMRTV}. Our method
 also works at non zero external spins .

We have written still a more compact 
representation of the GFT with EPR/FK propagator in terms of traces
\cite{BGR}.

Finally let us mention recent results which establish the 1/N expansion for colored models
and prove that spheres dominate in all dimensions \cite{Gur3,GurRiv,Gur4}.
This is encouraging for generating a large topologically trivial world as we know it.
In colored theory the issue of respecting suitable discretized diffeomorphism invariance seems
also more promising \cite{BGO}.
  
\subsection{What remains to be done}

Essentially everything! 
\begin{itemize}

\item\medskip  We need to establish results such as the dominance
of spherical graphs not only in the BF but also in the EPR/FK colored models.
The jacket bounds of the $1/N$ expansion \cite{Gur3,GurRiv,Gur4} 
seem a promising starting point in this direction.

\item\medskip  We need to cut propagators with non trivial spectra
like the EPR/FK into {\emph{slices}}  according to their spectrum. 
Such a slice requires two cutoffs, one for the top one for the bottom of the slice.
Then we can perform a true multiscale RG analysis, study the ``high" subgraphs, 
which should obey the new locality principle, and establish their correct power counting. 
Only then can we find whether some of these models are renormalizable or not.

\item\medskip  We need to find out whether symmetries such as the tantalizing topological
$BF$ symmetry recovered at $\gamma =1$ create fixed points of the RG analysis.

\item\medskip  We need to get a better glimpse of whether a phase transition to a condensed
phase can lead to emergence of a large smooth space-time.

\end{itemize}
  
\section{Conclusion and further Outlook}

Let us warn that this section is of a purely speculative nature and that
of course none of my collaborators are responsible for these speculations which will turn out to be wrong! 

Suppose some group field theory with the simple pentachore vertex and a yet to be specified
propagator turns out to be renormalizable with an ultraviolet fixed point and a low energy 
phase transition leading to the ordinary effective world of general relativity on a four dimensional space-time.
What would it mean for physics?

The physics of the universe from far ``before" the big bang to the current cosmos could be summarized
as a single giant renormalization group (RG) trajectory. It would flow out from very near a simple
very high or infinitely high ``transplanckian" fixed point\footnote{Of course the next mystery would be to understand
the vicinity of this fixed point and why the world emerged at all out of it...}. 
This fixed point might be characterized by the value 1 of the 
Immirzi parameter and might be topological in nature.  Our world could then be called asymptotically topological,
just like Yang-Mills theory is asymptotically free.
We hope the vicinity of that transplanckian fixed point could be fully understood with analytical tools which do not
require computer simulations, like is the case for integrable systems and probably for the Grosse-Wulkenhaar fixed point \cite{GW2}.
Cooling to the big bang and below, the RG trajectory of the universe could undergo many phase transitions 
until our large and smooth world appears to us as it is today. The emergence of space time, called geometrogenesis, would occur through a condensation of the vertices of group field theory
into a non perturbative phase. Through later transitions space-time cools further and becomes filled with the particles
and fields of the standard model at small scales  and with the galaxies we actually observe at large scales. It is only after the geometrogenesis that
we can really distinguish between these large infrared and small ultraviolet scales in the traditional sense; before geometrogenesis
there are only the abstract ``ultraspin" scales. Ordinary distances cannot be measured and time itself 
should be replaced by this abstract notion of ``scale". We could rather imagine the world as made of loosely interacting pentachores, sort of atoms of space time
that would float in an abstract vacuum which is no space-time at all. These atoms have not condensed yet,
that is they are not bound into any specific manifold.

This transplanckian era of the universe would be
characterized by the flow of the Immirzi parameter from 1 to some non trivial value.  
Slightly below the Planck scale and geometrogenesis, semiclassical computations involving classical geometry coupled to
quantized matter, such as the computation of the Bekenstein-Hawking radiation, would begin to make sense, 
as macroscopic areas and horizons start to appear. Therefore 
the Immirzi parameter, intrinsically quantum geometric, could freeze
there, at a value which we might fix
so as to recover the usual Bekenstein-Hawking entropy. This would give a renormalized
effective value for $\gamma$ which in loop quantum gravity
is usually considered to be close to $\log 3/ \pi \sqrt 2 $ \cite{Dreyer}.

However summarizing the transplanckian physics as a simple flow of $\gamma$
would not be quite right yet. We also expect the coupling constant $\lambda$ of the
group field theory vertex to move, and probably a few other operators as well, which might be related to the future matter and radiation fields of the standard model.
In the model case of the 
Grosse-Wulkenhaar model, fixing the wave function factor in front of the $p^2$ term to 1
is a way of fixing the energy scales of the renormalization group. Then two other
parameters have non-trivial logarithmic RG flows, namely $\Omega$, the harmonic potential term which measures the rate of mixing of infrared and ultraviolet, and 
the coupling constant $\lambda$. At $\Omega =1$, hence perfect mixing, the beta function vanishes because of the enhanced symmetry \cite{beta1}. 
By analogy we expect also both $\gamma$ and $\lambda$ to flow from a fixed point at $\gamma=1$, at which the beta function should vanish
because of the purely topological nature of the GFT at that point.

Running the RG flow from that $\gamma =1$ fixed point towards our effective world,
the coupling constant might grow from a small value to a larger
value where a non-perturbative phase transition occurs. 
This is what happens in the BCS theory of superconductivity or in QCD. 
For this condensation to occur and some manifold-like space time
to emerge\footnote{We would like also this manifold or at least its main four-dimensional part to be topologically trivial, i.e. spherical, 
because our four dimesnional universe does not look full of macroscopic holes or handles.},
 two main things are required. First
perturbation theory should be at the verge of diverging so that pentachores
should pullulate; second they should energetically prefer to glue in a nice and coherent pattern so as to create a smooth manifold, say a sphere.
Both phenomena should occur when $\lambda$ reaches a critical value $\lambda_c$
at which the sum of all `spherical graphs" would diverge. A spherical graph is the analog
of a planar graph in matrix models. Just like planar graphs dominate in
power counting in matrix and in Grosse-Wulkenhaar models, spherical graphs dominate in colored 4D GFTs \cite{GurRiv}. 
Just like the asymptotic behavior of the planar series at large order $n$ is in $K_P^n$ 
we expect the asymptotic behavior of these  spherical graphs to be in $K_S^n$ at large order, with
an analytically computable value of $K_S$. The value of $\lambda_c$ should simply
be $1/K_S$.

The vicinity of the phase transition being around $\lambda = \lambda_c$
might perhaps be probed non perturbatively with a double scaling limit. In any case
this phase transition should be more complex than BCS condensation which is vector like, or
quark confinement which is matrix like; it is ``4d-manifold-like" hence a new step in complexity. 
This part of the scenario is quite murky and it may be that for some time we have to 
investigate it through computer simulations, just like quark confinement 
and hadron formation are still studied today.

\medskip
\noindent{\bf Acknowledgments} 
This review benefitted from discussions with many people.
Let me thank more particularly Razvan Gurau and Jacques Magnen, and also Joseph Ben Geloun, Harald Grosse, Thomas Krajewski, 
Karim Noui, Daniele Oriti, Carlo Rovelli, Matteo Smerlak, Simone Speziale, 
Adrian Tanasa, Patrizia Vitale and Raimar Wulkenhaar.

\end{document}